# Study of the microwave electrodynamic response of $MgB_2$ thin films

A. Andreone[a], A. Cassinese[a], C. Cantoni[b], E. Di Gennaro[a], G. Lamura[a], M. G. Maglione[a], M. Paranthaman[b],

M. Salluzzo[a], R. Vaglio[a]

[a] I.N.F.M. and Dipartimento di Scienze Fisiche, Universita' degli Studi *Federico II*, Napoli, ITALY

[b] Oak Ridge National Laboratory, Oak Ridge, Tennessee, U.S.A.

**Abstract.** We present a study on the power dependence of the microwave surface impedance in thin films of the novel superconductor $MgB_2$. 500 nm thick samples exhibiting critical temperatures ranging between 26 and 38 K are synthesized by an *ex-situ* post-anneal of e-beam evaporated boron in the presence of an Mg vapor at 900 °C. Preliminary results on films grown *in situ* by a high rate magnetron sputtering technique from stoichiometric $MgB_2$ and Mg targets are also reported. Microwave measurements have been carried out employing a dielectrically loaded niobium superconducting cavity operating at 19.8 GHz and 4 K. The study shows that the electrodynamic response of $MgB_2$ films is presently dominated by extrinsic sources of dissipation, appearing already at low microwave power, likely to be ascribed to the presence of grain boundaries and normal inclusions in the samples.

**Corresponding author:** Antonello Andreone, Universita' degli Studi di Napoli *Federico II*, Facolta' di Ingegneria, Piazzale Tecchio 80, I-80125 Napoli, ITALY
Fax: ++39-081-2391821
E-mail: andreone@unina.it

The electrodynamic response of a superconductor in the microwave region represents an important tool for the accurate determination of applied and fundamental parameters of the material under test. In the newly discovered $MgB_2$ compound [1], the search for the intrinsic and extrinsic mechanisms responsible for the microwave absorption presents also very promising technological implications. Its high critical temperature together with the very good metallic properties, and the potential easy of deposition on large areas due to its simple binary structure, makes $MgB_2$ an almost ideal candidate as substitute for Nb in superconducting accelerating cavities operating at 4 K. In wireless telecommunications, $MgB_2$ may exploit its ability to be grown on cheaper substrates (namely, sapphire or silicon), rather than on MgO and $LaAlO_3$, for the development of passive devices operating in cryocoolers at about 20 K

The surface impedance $Z_s = R_s + iX_s$ is surely the most relevant parameter for microwave applications. $R_s$ provides a quantitative estimate of the dissipation induced by the e.m. field at the sample surface, whereas the measurement of the imaginary part provides direct information on the magnetic penetration depth $\lambda$, since $X_s = \mu_0 \omega \lambda$ ($\omega$ is the angular frequency). We present here a study on the power dependence of $Z_s$ of $MgB_2$ samples in thin film form.

Films were obtained by appropriate post-annealing of an electron-beam evaporated B precursor (500 nm) grown directly on an r-plane oriented $Al_2O_3$ single-crystal $10\times10\times0.5$ $mm^3$ substrate [2]. After deposition the Boron film was subsequently sandwiched between cold pressed $MgB_2$ pellets, along with excess Mg turnings, packed inside a crimped Ta cylinder and finally annealed at 890 °C for 20-25 min. Best samples showed values of the critical temperature $T_c$ of about 39 K with transition widths $\Delta T_c$ (10-90%) of 0.1 K and a residual resistivity ratio $RRR$ of about 3 (Fig. 1). Typical θ-2θ scan revealed the presence of a c-axis aligned film whereas the pole figure showed a random in-plane texture. AFM study showed the presence of a polycrystalline-type surface with grains up to 100 nm of diameter.

$MgB_2$ thin films were also grown *in-situ* by a magnetron sputtering technique in a UHV system ($10^{-7}$ Pa) equipped with 3 focused 2" magnetron sources. The substrates (sapphire or MgO) were placed 'on axis' at 7 cm from the target surface on the platelet of a molybdenum heater that can operate up to 1200 °C under vacuum. Deposition runs were performed co-depositing $MgB_2$ and Mg from stoichiometric targets at equal sputtering power (500 W) for 10 min. on cold substrates, resulting in a Mg rich Mg-B precursor film. The film was then annealed at 600 °C for 10 min. in Mg flow from the Mg target operated at the same sputtering power. During heater warm-up (cool-down) the sample was removed from the heater using tweezers mounted on a wobble-stick. The resulting films were about 1 μm thick, with 100 nm surface roughness as measured by AFM. Resistive transition showed a maximum $T_c$ of 22 K and a transition width of about 1 K. The residual resistivity ratio was 1.1 for the best sample.

Selected microwave measurements were performed on a number of samples grown by e-beam evaporation, having critical temperature ranging between 26 and 38 K.

The surface impedance was investigated by means of a dielectrically loaded resonator, consisting of a high purity ($RRR > 500$) niobium cylindrical shield of diameter 9.5 mm short-circuited at both ends by two plates. A cylindrical dielectric sapphire rod of diameter 7 mm and height 3.5 mm is placed and centered between the two parallel plates. The $TE_{011}$ field of the resonator is excited and detected by two semi-rigid coaxial cables, each having a small loop at the end, and the resonant frequency $f$ and $Q$-factor are measured in the transmission mode.

For the field measurements of the surface impedance on $MgB_2$ films we increased the input power $P_{in}$ feeding the cavity from – 40 to + 10 dBm at liquid helium temperature and replacing one end with the sample under test.

In fig. 1 the surface resistance $R_s$ versus the circulating power $P_c = Q_0 P_{in} (1-S_{11}^2 - S_{12}^2)$, where $Q_0$ is the unloaded quality factor and $S_{11}$ and $S_{12}$ are the measured scattering parameters of the cavity, is shown. At very low microwave power all samples show the same feature, characterized by a rapid increase of the surface resistance. In high temperature superconductors this has been commonly interpreted as flux flow losses caused by trapped flux in weak links [3]. However, this is somewhat unlikely in $MgB_2$, where magnetization [4] and non resonant microwave absorption studies on bulk samples [5] indicate negligible effects of intergranular weak links. Increasing the input power, there is a dramatic change in the dependence, since vortices start to enter the material. Due to the high frequency of operation, flux pinning is not encountered, and vortices can flow relatively easily. Results show that films having a reduced value of the critical temperature have also the lowest values of the surface resistance. This is not really surprising since the residual surface resistance is often an unpredictable function of the quality of the sample surface. This is especially true in the case of the novel superconducting $MgB_2$ material, where the procedure for thin film growth is still quite irreproducible. A comparison between our surface resistance measurements and data that have appeared in recent literature is rather encouraging. One can see that these films show a surface resistance of 2 mΩ at 19.8 GHz, which is significantly lower than previous reports, scaled at 19.8 GHz using the ordinary $\omega^2$ scaling law, i.e. 3-4 mΩ for films [6, 7], 10 mΩ for pellets [8] and is only slightly larger than the result obtained on wires [8] which show the lowest surface resistance of about 0.7 mΩ.

Fig. 2 shows the change in the resonant frequency $\Delta f = (f - f_{min})/f_{min}$, where $f_{min}$ is the value at the minimum input power, versus $P_c$. For small variations, the relation $\Delta f \propto \Delta\lambda$ holds, therefore this is equivalent to look at the surface reactance changes. While this quantity displays the same dependence as $R_s$ in film #1 and #2, its behavior is markedly different in film #3. In this latter sample, non linearity in the inductance is the primary source of microwave losses. This is also confirmed by the analysis of the dimensionless parameter $r = \Delta R_s/\Delta X_s = \Delta(1/Q)/\Delta f/f_{min})$, that has been widely used in different resonators to identify the source of nonlinearity in superconducting materials since it is independent of any geometrical factor [9, 10]. $r$ for the three samples under investigation is reported in table 1, together with the main transport properties, and it is of the order of unity for film #1 and #2, whereas it is one order of magnitude smaller for film #3. These distinctive features indicate that nonlinearity in the electrodynamic response is caused by different mechanisms, likely flux flow losses in film #1 and #2 and penetration of vortices in weakly coupled grains in film #3 [10].

In conclusion, a comparison of surface impedance versus power data amongst films of the recently discovered $MgB_2$ compound shows that the source of microwave loss can be markedly different, depending on the transport and structural properties of the samples. In particular, it is found that films exhibiting the lowest critical temperature (~ 26 K) show also the smallest residual losses at 20 GHz.

References.

Figure captions.

Figure 1: resistivity vs temperature for an $MgB_2$ film grown by the *ex situ* post-annealing e-beam technique

Figure 2: the surface resistance $R_s$ as a function of the power $P_c$ circulating in the superconducting cavity at 4 K: film #1 (▲), film #2 (■), film #3 (●)

Figure 3: the normalised change in the resonant frequency $\Delta f$ as a function of the power $P_c$ circulating in the superconducting cavity at 4 K: film #1 (▲), film #2 (■), film #3 (●)

Table 1: the main transport parameters measured for the films under study.

| Sample | $T_c$(K) | $J_c$(MA/cm$^2$) @ 4 K | $\rho_{300}$(μΩ·cm) | $\lambda(0)$ (μm) | $R_{res}$(mΩ) | $r$ |
|---|---|---|---|---|---|---|
| Film #1 | 26.0 | 0.4 | 900 | 1.2 | 0.4 | 1 |
| Film #2 | 26.5 | 0.3 | - | 1.0 | 0.3 | 2 |
| Film #3 | 37.9 | 6.4 | 22 | 0.1 | 1.7 | 0.01 |

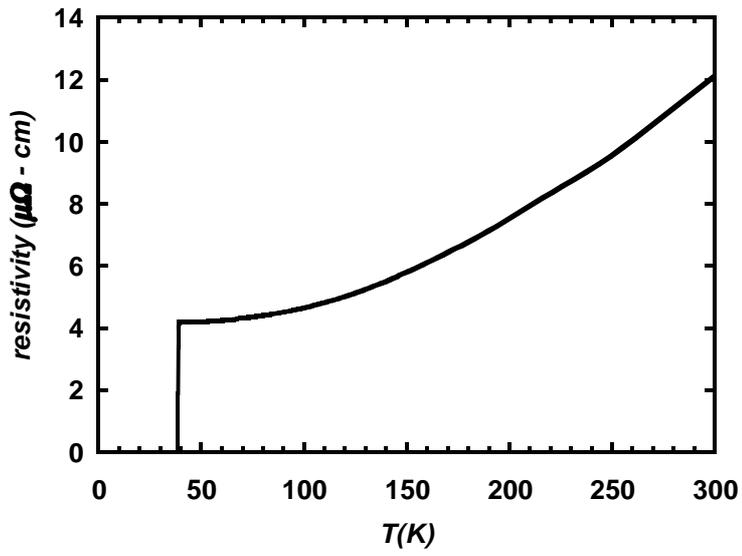

fig. 1     A. Andreone *et al.* (N1.2-25)

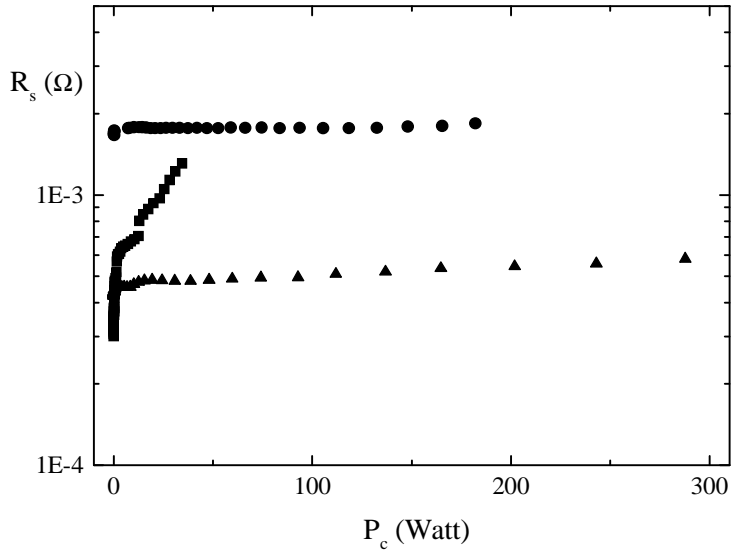

fig. 2    A. Andreone *et al.* (N1.2-25)

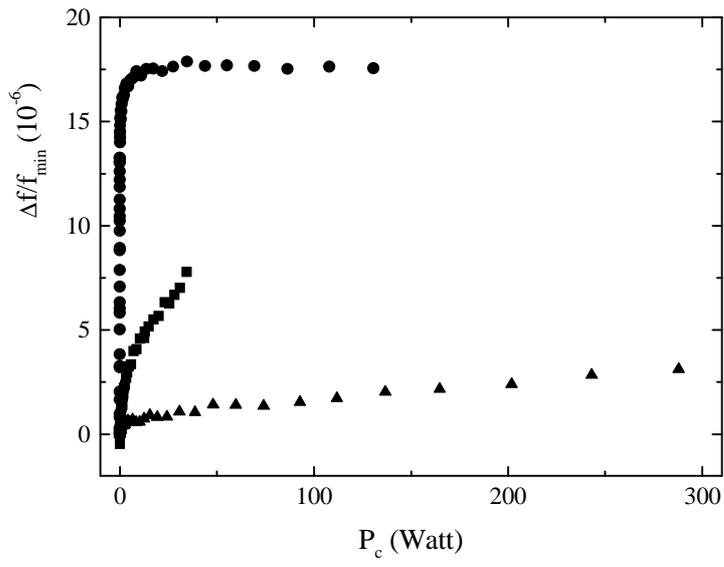

fig. 3   A. Andreone *et al.* (N1.2-25),